\documentclass[twocolumn,showpacs,preprintnumbers,amsmath,amssymb]{revtex4}

\usepackage{graphicx}
\usepackage{dcolumn}
\usepackage{bm}

\begin{document}


\title { Collapsing transition of spherical tethered surfaces with many holes}

\author{Hiroshi Koibuchi}
 \email{koibuchi@mech.ibaraki-ct.ac.jp}
\affiliation{%
Department of Mechanical and Systems Engineering, Ibaraki National College of Technology, Nakane 866 Hitachinaka, Ibaraki 312-8508, Japan}%

\date{\today}

\begin{abstract}
We investigate a tethered (i.e. fixed connectivity) surface model on spherical surfaces with many holes by using the canonical Monte Carlo simulations. Our result in this paper reveals that the model has only a collapsing transition at finite bending rigidity, where no surface fluctuation transition can be seen. The first-order collapsing transition separates the smooth phase from the collapsed phase. Both smooth and collapsed phases are characterized by Hausdorff dimension $H\!\simeq\!2$, consequently, the surface becomes smooth in both phases. The difference between these two phases can be seen only in the size of surface. This is consistent with the fact that we can see no surface fluctuation transition at the collapsing transition point. These two types of transitions are well known to occur at the same transition point in the conventional surface models defined on the fixed connectivity surfaces without holes. 
\end{abstract}

\maketitle
\section{Introduction}\label{intro}
Two-dimensional surfaces emerge as interfaces between two different materials. Biological membranes are considered as such two-dimensional interfaces separating some biological materials. The model of Helfrich, Kleinert, and Polyakov (HPK) \cite{HELFRICH-1973,POLYAKOV-NPB1986,KLEINERT-PLB1986} describes the mechanics of such surfaces on the basis of the two-dimensional differential geometry and the statistical mechanics \cite{NELSON-SMMS2004,Gompper-Schick-PTC-1994,Bowick-PREP2001}. 

One interesting topic is the crumpling transition, which has long been investigated theoretically \cite{Peliti-Leibler-PRL1985,DavidGuitter-EPL1988,PKN-PRL1988}
 and numerically \cite{KANTOR-NELSON-PRA1987,KD-PRE2002,KOIB-PRE-2005,KOIB-NPB-2005,Baum-Ho-PRA1990,CATTERALL-NPBSUP1991,AMBJORN-NPB1993,KOIB-EPJB-2005}. The surface fluctuations grow anomalously at certain finite bending rigidity in the surface models, where a surface collapsing phenomena can also be seen at the same transition point. These two phenomena have been understood as the crumpling transition. The surface fluctuation of the model is characterized by fluctuations of the bending energy $S_2$ and, consequently, we can see an anomalous peak in the corresponding specific heat $C_{S_2}$. On the other hand, the collapsing phenomena are characterized by a discontinuous change of the mean square size $X^2$ and, then, we can see a discontinuous (or a continuous) change in the roughness exponent or in the Hausdorff dimension.

We have not yet seen that any surface models undergo only one of the two transitions. Experimentally, it was reported that a surface fluctuation transition is accompanied by a collapsing transition in certain artificial membranes \cite{CNE-PRL-2006}. However, it is possible to consider that these phenomena are two different ones. Buckling of a thin elastic shell is one of the collapsing phenomena \cite{PA-PRE2003}, and it is not always accompanied by the surface fluctuations. In biological membranes, the surface fluctuation called a rippling transition \cite{LL-PRB1987} is not always accompanied by the surface collapsing phenomena.   

In this paper, we show that the collapsing transition occurs in a fixed connectivity surface model at finite bending rigidity $b_c$, where no surface fluctuation transition is observed. The surface model is defined by the standard HPK Hamiltonian, which is discretized on triangulated spherical surfaces with many holes. The starting surface configuration for the Monte Carlo (MC) simulations is constructed such that the ratio of the area of the holes to that of the surface including the holes is fixed in the thermodynamic limit. 

 The transition point $b_c$($=\!1.4\!\sim\!1.5$) is relatively larger than that ($b_c\!\simeq\!0.77$) in the same model on spherical surfaces without the holes. As a consequence, the surfaces with holes are relatively smooth at the collapsing transition point $b_c$. Then, we have $H\!\simeq\!2$ even in the collapsed phase. 

The result in this paper indicate a possible collapsing transition which is not accompanied by the surface fluctuation transition in biological or artificial membranes, although the self-avoiding property \cite{GREST-JPIF1991,BOWICK-TRAVESSET-EPJE2001,BCTT-PRL2001} is not assumed in the model. A crumpling phenomenon can also be seen on thin elastic sheets and has been investigated by singularity analysis \cite{CM-PRL1998,SCM-SCIE2000}. However, the collapsing transition in biological membranes is not yet completely understood because of the complexity of biological membranes. 
 
\section{Model and Monte Carlo technique}\label{model}
In order to construct the lattice that has holes, we start with the icosahedron. Firstly, we divide the edges of the icosahedron into $\ell$ pieces of the uniform length $a$. Then, we have a triangulated surface of size $N_0\!=\!10\ell^2\!+\!2$, which is the total number of vertices in the triangulated sphere without holes. Then an edge length of the icosahedron corresponds to $\ell$ edges in the triangulated sphere. Secondly, the $\ell$ edges are divided into $m$ groups ($m\!=\!1,2,\cdots$), where each group has length $L\!=\!\ell /m$ in the unit of $a$ if $m$ divides $\ell$. As a consequence, we have a sublattice of edge length $La$ in the triangulated sphere of size $N_0$. We should note that this sublattice makes compartments on the lattice. Finally, one part of the compartments in the sublattice is labeled as holes, and the other remaining part is labeled as the lattice points. Thus, we have a triangulated lattice with many holes. The total number of holes in one face of the icosahedron is given by $m(m-1)/2$ and, hence the total number of holes is $10m(m-1)$ on the lattice. Because of the holes on the surface, the total number of vertices $N$ of the lattice with holes are reduced from $N_0$, which is the total number of the original triangulated lattice as stated above. We note that $N$ includes the vertices on the boundary of holes.  The size of lattice can be characterized by $(N,m)$.

\begin{figure}[htb]
\includegraphics[width=8.5cm]{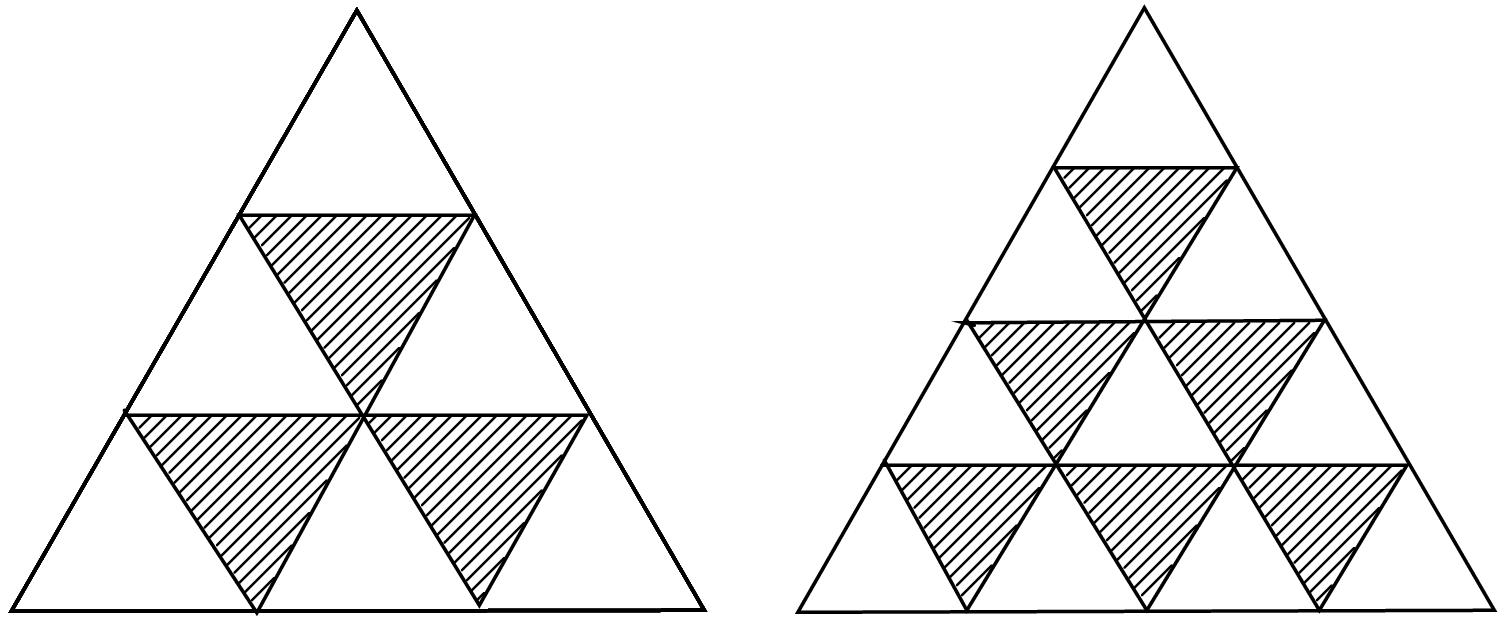}
\includegraphics[width=8.5cm]{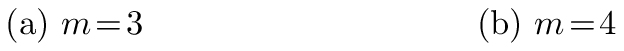}
\caption{A face of icosahedron corresponding to (a) $m\!=\!3$ and (b) $m\!=\!4$.  The shaded triangles excluding their corners correspond to the holes.  The total number of holes in the face is (a) $3$ and (b) $6$, which are given by $m(m-1)/2$.} 
\label{fig-1}
\end{figure}
We show the holes in a face of the icosahedron in Figs.\ref{fig-1}(a) and \ref{fig-1}(b), which respectively correspond to $m\!=\!3$ and $m\!=\!4$. Note that the faces are those of the triangulated surfaces, whose total number of vertices is given by the above described number $N_0\!=\!10\ell^2\!+\!2$. The small triangles including the shaded ones in the figures form the above-mentioned sublattice or compartments and, therefore, they are triangulated with more fine triangles. At the corners of each shaded triangle, the three fine triangles, which are not shown in the figures, are excluded from the hole (or equivalently included in the surface).  

\begin{figure}[htb]
\includegraphics[width=8.5cm]{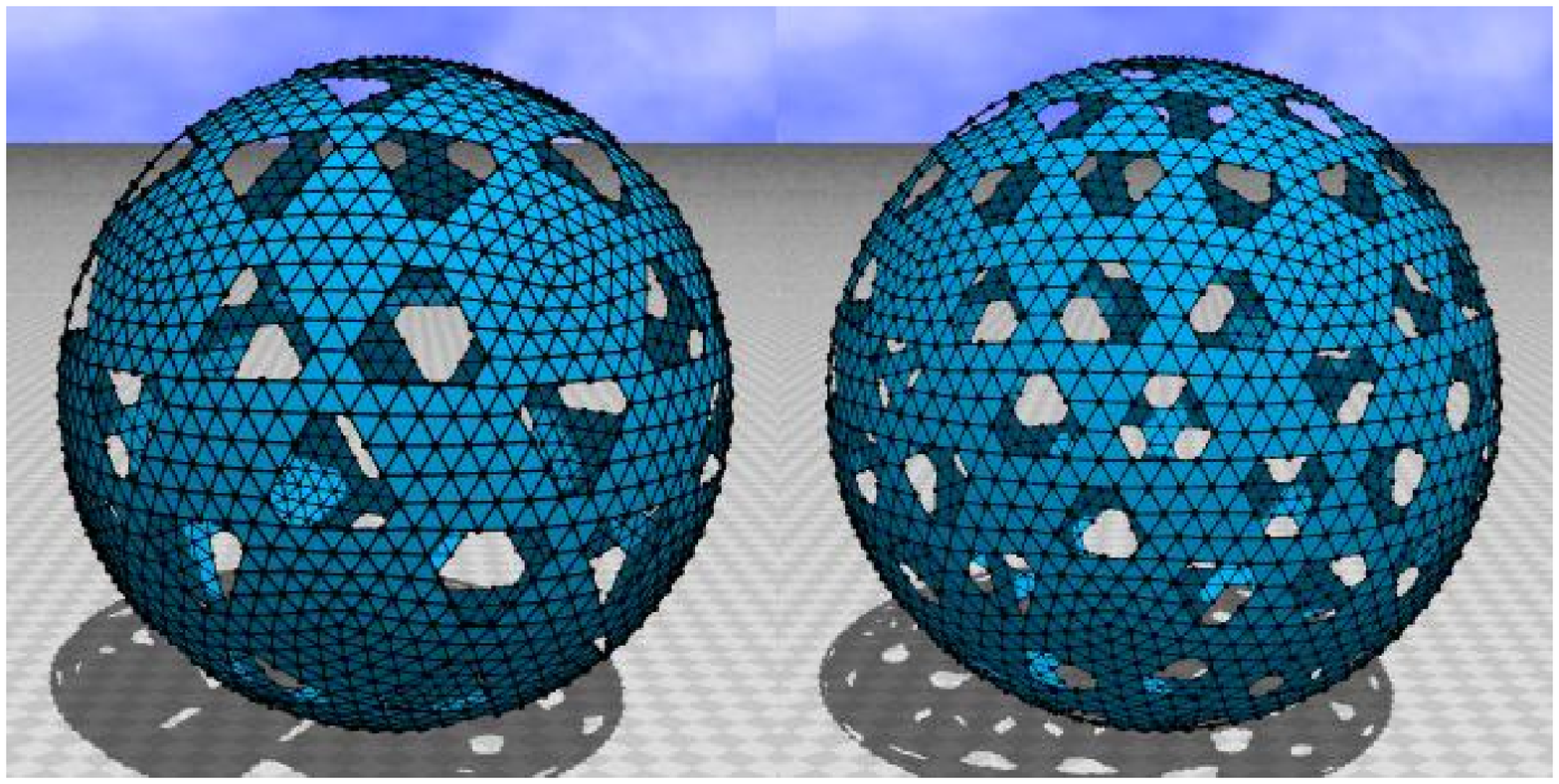}
\includegraphics[width=8.5cm]{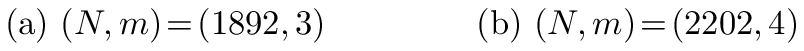}
\caption{(Color online) Starting configuration of surfaces of (a) $(N,m)\!=\!(1892,3)$ and  (b) $(N,m)\!=\!(2202,4)$. The total number of holes is  $60$ in (a), and it is $120$ in (b).} 
\label{fig-2}
\end{figure}
Figures \ref{fig-2}(a) and \ref{fig-2}(b) are the lattices of $(N,m)\!=\!(1892,3)$, and $(N,m)\!=\!(2022,4)$. We note that the ratio $R_m(N)$ of the area of holes to that of the surface including the holes is $R_m(N)\!\to\!(m-1)/2m$ in the limit of $N\to \infty$ or equivalently $N_0\to \infty$. In fact, the total number of triangles in a hole is $L^2\!-\!3$ and, then, the total number of triangles in the holes is $10m(m\!-\!1)(L^2\!-\!3)$, which is easily understood since the total number of faces in the icosahedron is $20$, and the total number of holes in a face is $m(m\!-\!1)/2$ as stated above. On the other hand, the total number of triangles on the triangulated sphere is $2N_0\!-\!4$. Then, we have $R_m\!=\! 10m(m\!-\!1)(L^2\!-\!3)/(2N_0\!-\!4)$. By using $L^2\!=\!(\ell/m)^2$ and $\ell^2\!=\!(N_0\!-\!2)/10$, we have $R_m=(m-1)/2m$ in the limit of $N_0\to \infty$.

We assume two values for $m$ such that
\begin{equation}
\label{values-of-m}
m=3,\quad m=4,
\end{equation}
and four sizes of lattices are assumed for each $m$. Table \ref{table-1} shows the total number of vertices $N$ and $N_0$ for each $m$ and the corresponding $R_m(N)$, where $N_0$ is the total number of vertices of the original lattice, which has no holes. We find that the ratio $R_m(N)$ in Table  \ref{table-1} are very close to the values $R_3(\infty)\!=\!1/3$ or $R_4(\infty)\!=\!3/8$, although $N$ is finite in Table \ref{table-1}.  
\begin{table}[hbt]
\caption{The total number of vertices $N$ and $N_0$ for each $m$, and the corresponding $R_m(N)$. $N_0$ is the total number of vertices of the original lattice, which has no holes.}
\label{table-1}
\begin{center}
 \begin{tabular}{ccccccc}
$m$  & $N$($N_0$) & $R_3(N)$ &$\quad$& $m$  & $N$($N_0$) & $R_4(N)$ \\
 \hline
  3  & 1892(2252)  & 0.3320  &$\quad$&  4   & 2202(2562)  & 0.3735   \\
  3  & 3512(4412)  & 0.3327  &$\quad$&  4   & 4402(3282)  & 0.3736  \\
  3  & 5612(7292)  & 0.3329  &$\quad$&  4   & 6042(7842)  & 0.3744  \\
  3  & 8192(10892) & 0.3330  &$\quad$&  4   & 7722(10242) & 0.3746 \\
 \hline
 \end{tabular} 
\end{center}
\end{table}

The model is defined by the partition function 
\begin{eqnarray} 
\label{Part-Func}
 Z = \int^\prime \prod _{i=1}^{N} d X_i \exp\left[-S(X)\right],\\  
 S(X)=S_1 + b S_2, \nonumber
\end{eqnarray} 
where $b$ is the bending rigidity, $\int^\prime$ denotes that the center of the surface is fixed in the integration. $S(X)$ denotes that the Hamiltonian $S$ depends on the position variables $X$ of the vertices. The Gaussian bond potential $S_1$ and the bending energy $S_2$ are defined by
\begin{equation}
\label{Disc-Eneg} 
S_1=\sum_{(ij)} \left(X_i-X_j\right)^2,\quad S_2=\sum_{(ij)} (1-{\bf n}_i \cdot {\bf n}_j),
\end{equation}
where $\sum_{(ij)}$ in $S_1$ is the sum over bonds $(ij)$ connecting the vertices $i$ and $j$ including those on the boundary, and $\sum_{(ij)}$ in $S_2$ is also the sum over bonds $(ij)$, which are edges of the triangles $i$ and $j$.  ${\bf n}_i$ in Eq. (\ref{Disc-Eneg}) is the unit normal vector of the triangle $i$. We note that $S_2$ is a quantity that is defined on the bonds.  Note also that $S_2$ is not defined on the boundary bonds, because ${\bf n}_i$ is not defined inside the holes. 

The unit of $b$ is $kT$, where $k$ is the Boltzmann constant and $T$ is the temperature. Let $\lambda$ be the string tension coefficient such that $S\!=\!\lambda S_1\!+\!bS_2$, then the length unit $a$ of the model is given by $\sqrt{kT/\lambda}$. Consequently, the unit of $S_1$ can be expressed by $a^2$ or $kT/\lambda$. Note also that we can arbitrarily choose the length unit $a$ because of the scale invariant property of $Z$ in Eq.(\ref{Part-Func}). Therefore, $\lambda\!=\!1$ was assumed in the simulations.

The total number of Monte Carlo sweeps (MCS) after the thermalization MCS are about $5\sim 6\times10^8$ for $(N,m)\!=\!(8192,3)$, $3\sim 4\times10^8$ for $(N,m)\!=\!(5612,3)$, and $2\sim 3\times10^8$ for $(N,m)\!\leq\!(3512,3)$ at $b$ close to the transition point. Relatively smaller number of MCS was performed at non-transition points. In the case $m\!=\!4$, the total number of MCS after the thermalization MCS are $5\sim 6\times10^8$ for $(N,m)\!=\!(7722,4)$, $3\sim 4\times10^8$ for $(N,m)\!=\!(6042,4)$, and $2\sim 3\times10^8$ for $(N,m)\!\leq\!(3282,4)$.

We use the canonical MC technique to simulate the integrations in the partition function. The vertices $X$ are shifted so that $X^\prime \!=\! X\!+\!\delta X$, where $\delta X$ is randomly chosen in a small sphere. The new position $X^\prime$ is accepted with the probability ${\rm Min}[1,\exp(-\Delta S)]$, where $\Delta S\!=\! S({\rm new})\!-\!S({\rm old})$. 

The radius of the small sphere for $\delta X$ is chosen so that the rate of acceptance for $X$ is about $50\%$. We introduce the lower bound $1\times 10^{-8}$ for the area of triangles. No lower bound is imposed on the bond length. 

\section{Results}\label{Results}
\begin{figure}[htb]
\includegraphics[width=8.5cm]{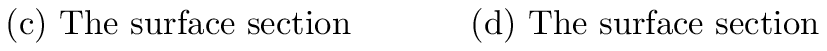}
\includegraphics[width=8.5cm]{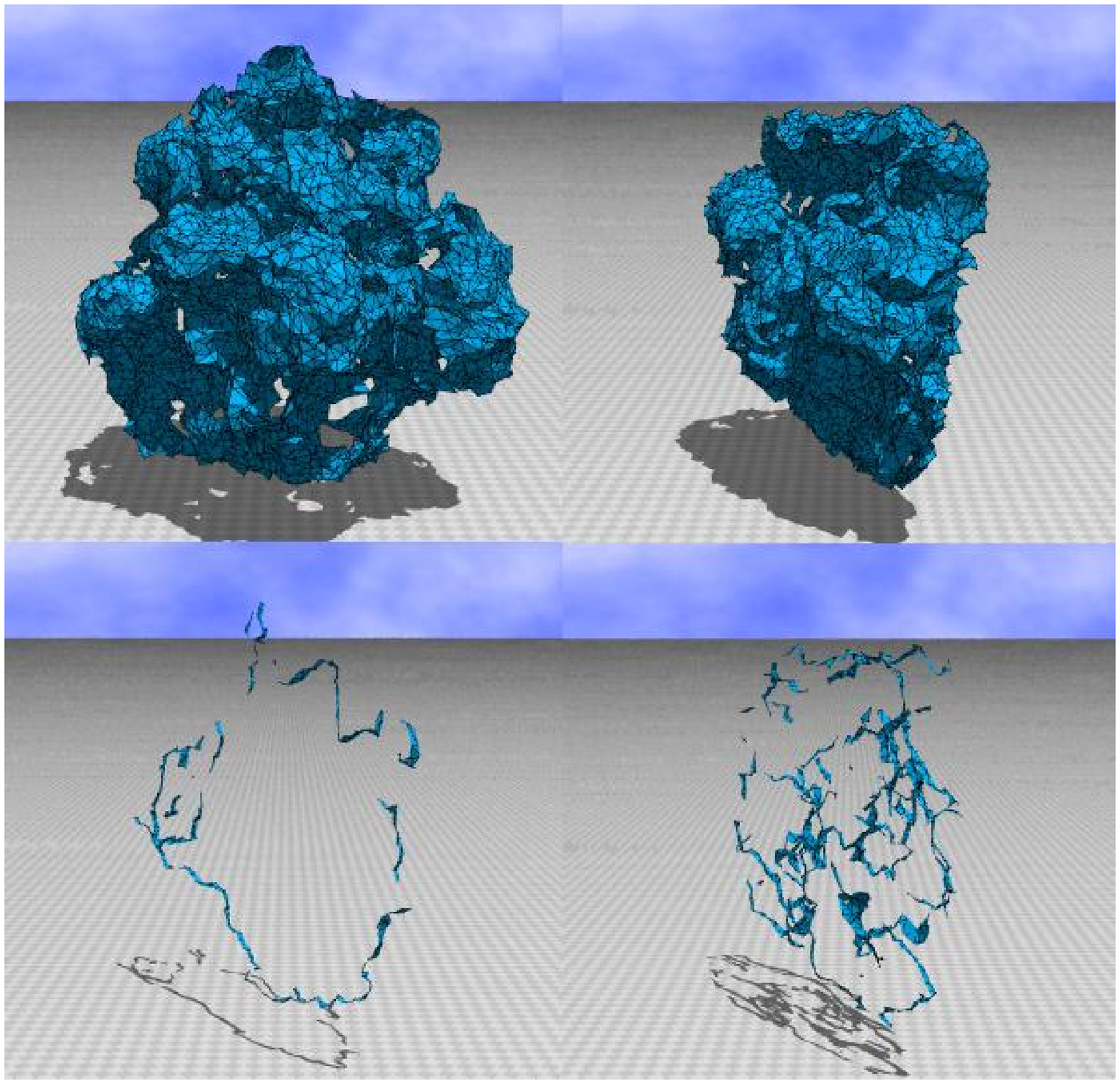}
\includegraphics[width=8.5cm]{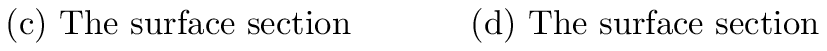}
\caption{(Color online) Snapshots of surfaces of $(N,m)\!=\!(8192,3)$ at $b\!=\!1.39$; (a) a smooth surface, (b) a collapsed surface, (c) the surface section of (a), and (d) the surface section of (b). The mean square size $X^2\!=\!127$ in (a), and $X^2\!=\!69$ in (b), where $X^2$ is defined by Eq.(\ref{X2}). Figures were drawn in the same scale.} 
\label{fig-3}
\end{figure}
Firstly, we show in Fig.\ref{fig-3}(a) a snapshots of the $(N,m)\!=\!(8192,3)$ surface, which is obtained in the smooth phase at the transition point $b\!=\!1.39$.  Figure \ref{fig-3}(b) shows a snapshot in the collapsed phase of the same surface at the same transition point. The surface sections of Figs.\ref{fig-3}(a) and \ref{fig-3}(b) are  shown in Figs.\ref{fig-3}(c) and \ref{fig-3}(d), respectively. We find from the surface sections that there is a lot of empty space inside the surface even in the collapsed surface. This implies that the surface is not completely collapsed in the collapsed phase. The vertices are not always confined in a small region; they are considered to be spread over some two-dimensional region in ${\bf R}^3$ such that the surface becomes almost smooth.  

\begin{figure}[htb]
\includegraphics[width=8.5cm]{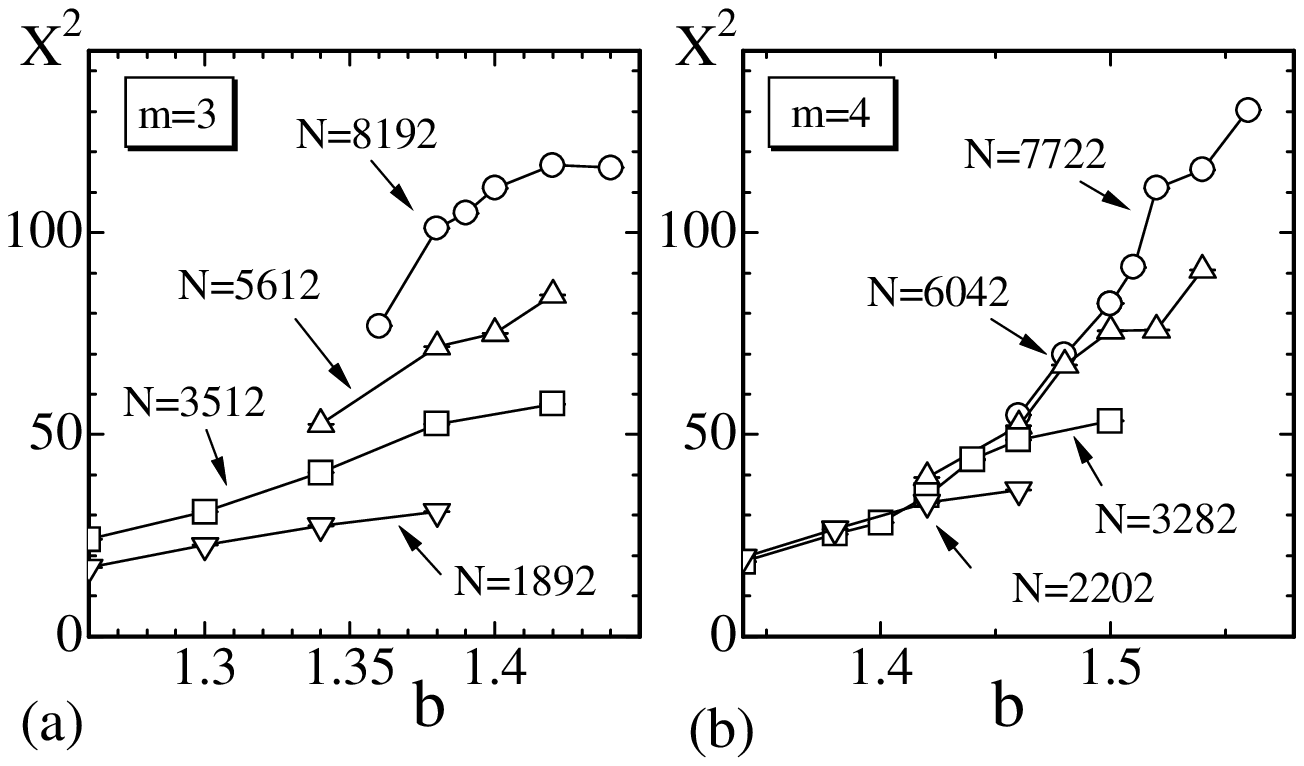}
\caption{The mean square size $X^2$ vs. $b$ obtained on the surfaces of (a) $m\!=\!3$, (b) $m\!=\!4$, and (c) $m\!=\!5$. } 
\label{fig-4}
\end{figure}
The mean square size $X^2$ is defined by
\begin{equation}
\label{X2}
X^2={1\over N} \sum_i \left(X_i-\bar X\right)^2, \quad \bar X={1\over N} \sum_i X_i,
\end{equation}
where $\bar X$ is the center of the surface. The mean square size of the snapshots in Fig.\ref{fig-3}(a) is $X^2\!=\!127$, whereas $X^2\!=\!69$ in Fig.\ref{fig-3}(b). These values of $X^2$ are typical to the smooth phase and to the collapsed phase on the $(N,m)\!=\!(8192,3)$ surface. 

Figures \ref{fig-4}(a) and \ref{fig-4}(b) show $X^2$ versus $b$ obtained on the surfaces of $m\!=\!3$ and $m\!=\!4$, respectively. The variation of $X^2$ versus $b$ seems to be not so sharp compared with the results of the tethered surface model in \cite{KOIB-PRE-2005}  even when $N$ becomes large. The surface size seems to change continuously in the model in this paper. 

\begin{figure}[htb]
\includegraphics[width=8.5cm]{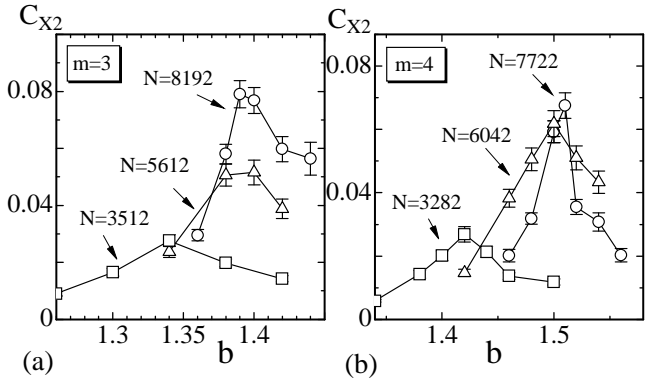}
\caption{The fluctuation $C_{X^2}$ vs. $b$ on the surfaces of (a) $m\!=\!3$ and (b) $m\!=\!4$. The error bars denote the statistical errors, which were obtained by the standard binning analysis.} 
\label{fig-5}
\end{figure}
The fluctuation $C_{X^2}$ of $X^2$ can be defined by 
\begin{equation}
\label{fluctuation-X2}
C_{X^2} = {1\over N} \langle \; \left( X^2 \!-\! \langle X^2 \rangle\right)^2\rangle,
\end{equation}
which is expected to reflect surface collapsing phenomena. Figures \ref{fig-5}(a) and \ref{fig-5}(b) are $C_{X^2}$ against $b$ obtained on the lattices of $m\!=\!3$ and $m\!=\!4$, respectively. Sharp peaks can be seen in $C_{X^2}$ and grow larger and larger when the size $N$ increases. This anomalous behavior of $C_{X^2}$ indicates a discontinuous change of the surface size. Therefore, this surface collapsing phenomenon can be viewed as a phase transition contrary to the expectation from the smooth variation of $X^2$ in Figs.\ref{fig-4}(a) and \ref{fig-4}(b). 

From Figs.\ref{fig-5}(a) and \ref{fig-5}(b) we find that the transition points $b_c$ are about $b_c\!=\!1.4$ and $b_c\!=\!1.5$ in the cases $m\!=\!3$ and $m\!=\!4$, respectively. Both of them are relatively larger than the transition point $b_c\!\simeq\!0.77$ of the same model on the fixed connectivity surface without holes \cite{KOIB-PRE-2005}.

\begin{figure}[htb]
\includegraphics[width=8.5cm]{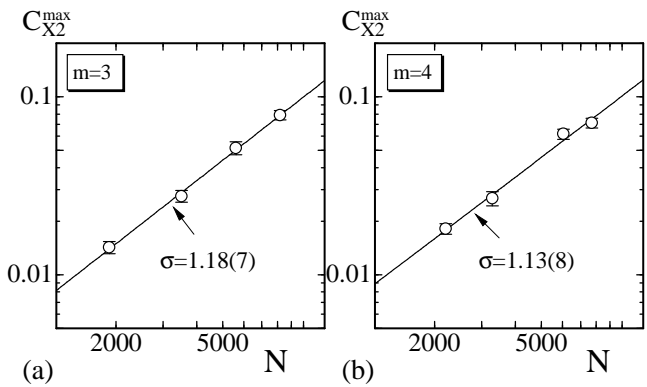}
\caption{Log-log plots of the peak values $C_{X^2}^{\rm max}$ against $N$, which were obtained on the surfaces of (a) $m\!=\!3$ and (b) $m\!=\!4$. The straight lines drawn in the figures were obtained by fitting the data to Eq.(\ref{scaling-CX2}).} 
\label{fig-6}
\end{figure}
In order to see the order of the transition, we plot the peak values $C_{X^2}^{\rm max}$ against $N$ in Figs.\ref{fig-6}(a) and \ref{fig-6}(b) in a log-log scale. The straight lines drawn in the figures were obtained by fitting the data to
\begin{equation}
\label{scaling-CX2}
C_{X^2}^{\rm max} \sim N^\sigma, 
\end{equation}
where $\sigma$ is a critical exponent of the transition. Then, we have
\begin{eqnarray}
\label{sigma-values}
&& \sigma = 1.18\pm0.07 \quad (m=3), \nonumber \\
&& \sigma = 1.13\pm0.08 \quad (m=4).  
\end{eqnarray}
These results indicate that the collapsing transition is of first order. In fact, we understand from the finite-size scaling theory \cite{PRIVMAN-WS-1989,BINDER-RPP-1997} that the maximum $C_{X^2}^{\rm max}$ should scale according to $N^\sigma (\sigma\!=\!1)$ at the transition point if the transition is of first order \cite{BNB-NPB-1993}.

\begin{figure}[htb]
\includegraphics[width=8.5cm]{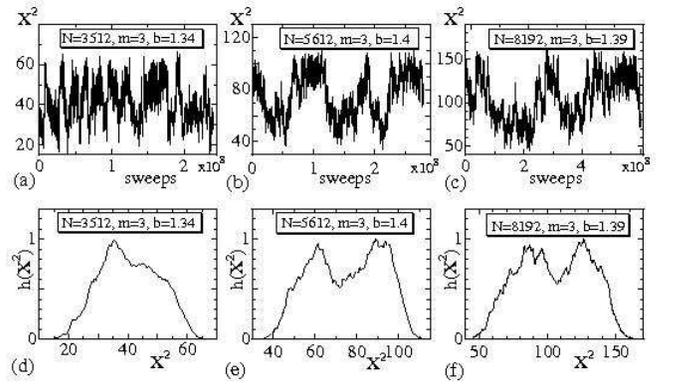}
\caption{The variation of $X^2$ against MCS obtained on the surfaces of (a) $(N,m)\!=\!(3512,3)$ at $b\!=\!1.34$, (b) $(N,m)\!=\!(5612,3)$ at $b\!=\!1.4$, and (c) $(N,m)\!=\!(8192,3)$ at $b\!=\!1.39$. The corresponding distribution $h(X^2)$ of $X^2$ are shown in (d), (e) and (f). } 
\label{fig-7}
\end{figure}
Figures \ref{fig-7}(a), \ref{fig-7}(b), and \ref{fig-7}(c) show the variation of $X^2$ versus MCS obtained at $b\!=\!1.34$ on the $(N,m)\!=\!(3512,3)$ surface,  $b\!=\!1.4$ on the $(N,m)\!=\!(5612,3)$ surface, and $b\!=\!1.39$ on the $(N,m)\!=\!(8192,3)$ surface, respectively. Figures \ref{fig-7}(d), \ref{fig-7}(e), and \ref{fig-7}(f) are the normalized distribution $h(X^2)$ (or histogram) of $X^2$ corresponding to the variations of $X^2$ in Figs.\ref{fig-7}(a), \ref{fig-7}(b), and \ref{fig-7}(c), respectively. We find a double peak structure in $h(X^2)$; one of the peaks represents the swollen and smooth state, which corresponds the snapshot in Fig.\ref{fig-3}(a), and the other represents the collapsed state, which corresponds the snapshot in Fig.\ref{fig-3}(b).   

\begin{figure}[htb]
\includegraphics[width=8.5cm]{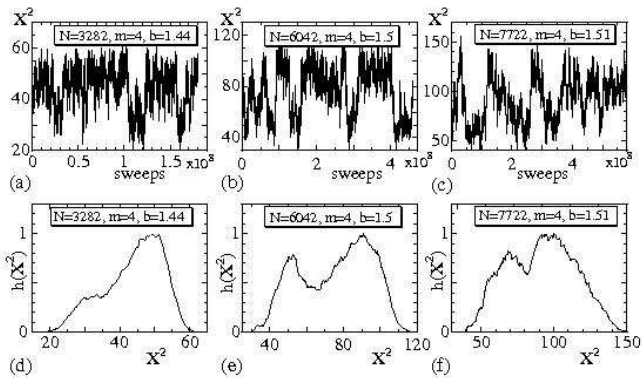}
\caption{The variation of $X^2$ against MCS obtained on the surfaces of (a) $(N,m)\!=\!(3282,4)$ at $b\!=\!1.44$, (b) $(N,m)\!=\!(6042,4)$ at $b\!=\!1.5$, and (c) $(N,m)\!=\!(7722,4)$ at $b\!=\!1.51$. The corresponding distribution $h(X^2)$ of $X^2$ are shown in (d), (e) and (f). } 
\label{fig-8}
\end{figure}
Figures \ref{fig-8}(a), \ref{fig-8}(b), and \ref{fig-8}(c) show the variation of $X^2$ versus MCS obtained at $b\!=\!1.44$ on the $(N,m)\!=\!(3282,4)$ surface,  $b\!=\!1.5$ on the $(N,m)\!=\!(6042,4)$ surface, and $b\!=\!1.51$ on the $(N,m)\!=\!(7722,4)$ surface, respectively. Figures \ref{fig-8}(d), \ref{fig-8}(e), and \ref{fig-8}(f) are the normalized histogram $h(X^2)$ of $X^2$ corresponding to the variations of $X^2$ in Figs.\ref{fig-8}(a), \ref{fig-8}(b), and \ref{fig-8}(c), respectively. The double peak structure can also be found in $h(X^2)$ in the case $m\!=\!4$. 

The double peaks in $h(X^2)$ allow us to estimate the Hausdorff dimension $H$ in the smooth state and that in the collapsed state, where $H$ is defined by
\begin{equation}
\label{Hausdorff-Dim}
X^2 \propto N^{2/H}.
\end{equation}
Note that we understand from the value of $H$ how smooth the surface is, because $H$ is defined so that $H$ varies depending on the distribution of vertices. In fact, we have $H\!\to\!2$ when the vertices form a two-dimensional region in ${\bf R}^3$, and we have $H\!\to\!3$ when the vertices form a three-dimensional region of constant density of vertices. Moreover, the vertices can form a zero-dimensional region (= a point) if no self-avoiding property is assumed, and then we have $H\!\to\!\infty$ in this case.

\begin{table}[hbt]
\caption{ The lower bound $X^{2 \;{\rm col}}_{\rm min}$ and the upper bound $X^{2 \;{\rm col}}_{\rm max}$ for obtaining the mean value $X^2({\rm col})$ in the collapsed state, and the lower bound $X^{2 \;{\rm smo}}_{\rm min}$ and the upper bound $X^{2 \;{\rm smo}}_{\rm max}$ for obtaining the mean value $X^2({\rm smo})$ in the smooth state. }
\label{table-2}
\begin{center}
 \begin{tabular}{cccccc}
$m$  & $N$ & $X^{2 \;{\rm col}}_{\rm min}$ & $X^{2 \;{\rm col}}_{\rm max}$ & $X^{2 \;{\rm smo}}_{\rm min}$ & $X^{2 \;{\rm smo}}_{\rm max}$   \\
 \hline
  3  & 1892 & 11 & 21  & 23 & 34  \\
  3  & 3512 & 21 & 39  & 42 & 65  \\
  3  & 5612 & 40 & 70  & 72 & 108  \\
  3  & 8192 & 50 & 100 & 110 & 160  \\
 \hline
  4  & 2202 & 12 & 24  & 26 & 39  \\
  4  & 3282 & 22 & 39  & 42 & 60  \\
  4  & 6042 & 35 & 70  & 75 & 118  \\
  4  & 7722 & 50 & 85  & 88 & 145  \\
 \hline
 \end{tabular} 
\end{center}
\end{table}
In order to obtain $X^2$ in the collapsed state, we evaluate the mean value $X^2({\rm col})$ by averaging $X^2$ in the range $X^{2 \;{\rm col}}_{\rm min}\!\leq\! X^2 \!\leq\! X^{2 \;{\rm col}}_{\rm max}$, where $X^{2 \;{\rm col}}_{\rm min}$ and $X^{2 \;{\rm col}}_{\rm max}$ are shown in Table \ref{table-2}.  The mean value $X^2({\rm smo})$ in the smooth state can also be evaluated by averaging $X^2$ in the range $X^{2 \;{\rm smo}}_{\rm min}\!\leq\! X^2 \!\leq\! X^{2 \;{\rm smo}}_{\rm max}$, where  $X^{2 \;{\rm smo}}_{\rm min}$ and $X^{2 \;{\rm smo}}_{\rm max}$ are also shown in Table \ref{table-2}. 

\begin{figure}[htb]
\includegraphics[width=8.5cm]{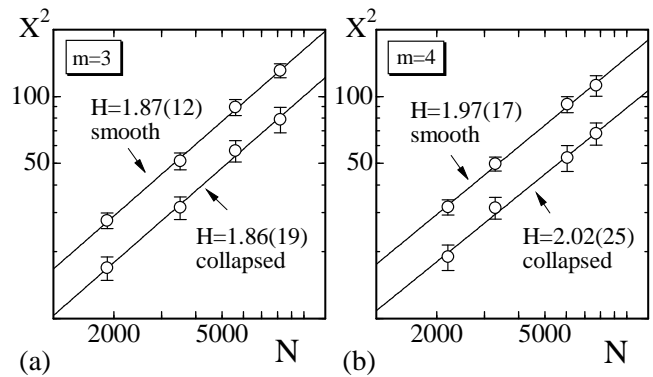}
\caption{Log-log plots of $X^2({\rm col})$ and $X^2({\rm smo})$ against $N$, which were obtained on the surfaces of (a) $m\!=\!3$ and (b) $m\!=\!4$. The straight lines drawn in the figures were obtained by fitting the data to Eq.(\ref{Hausdorff-Dim}).} 
\label{fig-9}
\end{figure}
The mean values  $X^2({\rm col})$ and $X^2({\rm smo})$ obtained on the surfaces $m\!=\!3$ and $m\!=\!4$ are respectively plotted in Figs.\ref{fig-9}(a) and \ref{fig-9}(b). The error bars denote the standard deviations, which were obtained by splitting the range of $X^2$ shown in Table \ref{table-2} into $10$ ranges; as a consequence, the errors become large (small) when the range is wide (narrow). The straight lines in the figures were drawn by fitting the data to Eq.(\ref{Hausdorff-Dim}) with the weight of inverse errors. Then, we have Hausdorff dimensions $H^{\rm col}$ and $H^{\rm smo}$ in the collapsed state and the smooth state such that
\begin{eqnarray}
\label{H-values}
H^{\rm col}= 1.86\pm 0.19, \;H^{\rm smo}= 1.87\pm 0.12 \;\; (m=3), \nonumber \\
H^{\rm col}= 1.97\pm 0.17,\; H^{\rm smo}= 2.02\pm 0.25 \;\; (m=4).
\end{eqnarray}
The values of $H$ in Eq.(\ref{H-values}) are all close to $H\!=\!2$, which is the topological dimension of the two-dimensional surfaces. Thus, we understand that not only the smooth state but also the collapsed state can be viewed as a smooth surface. Only difference between the smooth state and the collapsed state is in the size, which is characterized by $X^2$.  No difference can be seen in the Hausdorff dimensions of the smooth state and that of the collapsed state.

\begin{figure}[htb]
\includegraphics[width=8.5cm]{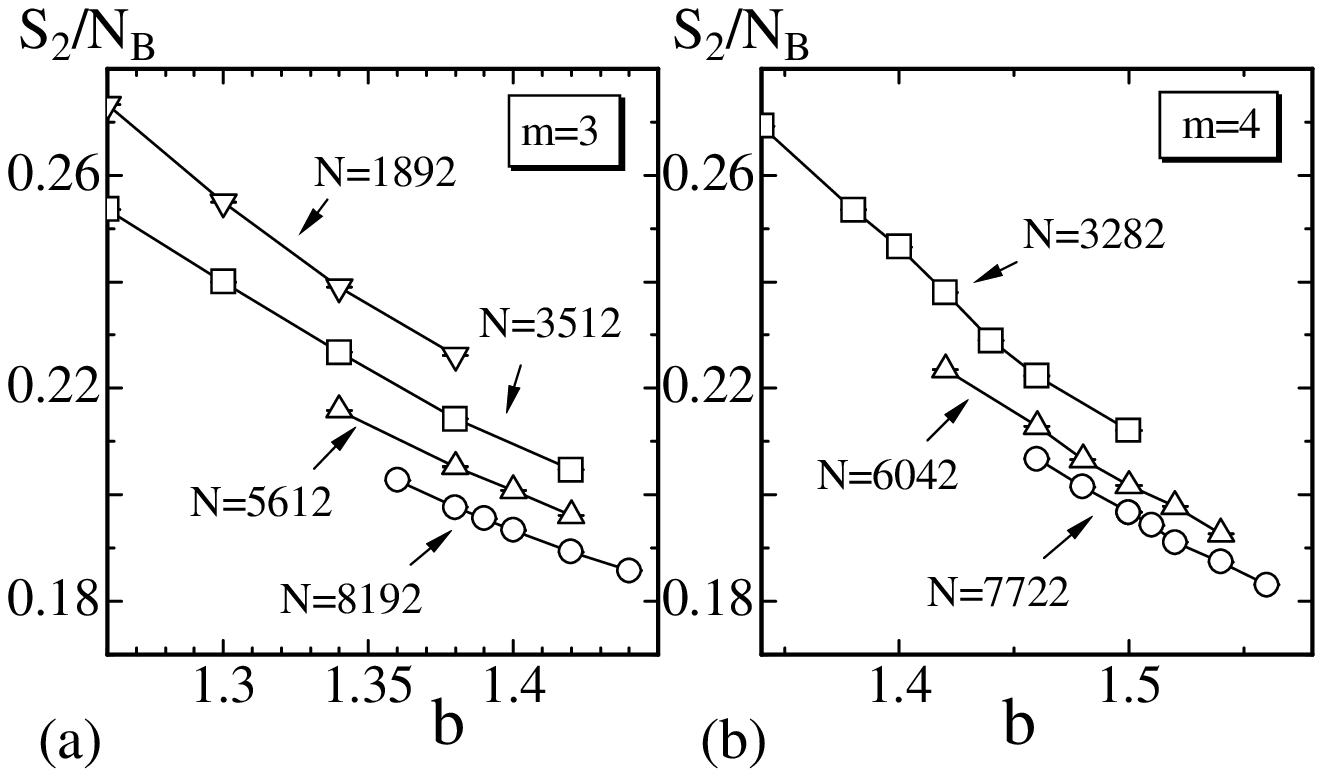}
\caption{The bending energy $S_2/N_B$ versus $b$ obtained on the surfaces of (a) $m\!=\!3$ and (b) $m\!=\!4$. $N_B$ is the total number of bonds. } 
\label{fig-10}
\end{figure}
Figures \ref{fig-10}(a) and \ref{fig-10}(b) show the bending energy $S_2/N_B$ versus $b$ obtained on the surfaces of $m\!=\!3$ and  $m\!=\!4$, respectively. The values of $S_2/N_B$ smoothly varies against $b$ close to the transition point even when $N$ is increased. This indicates that there is no surface fluctuation transition.  

\begin{figure}[htb]
\includegraphics[width=8.5cm]{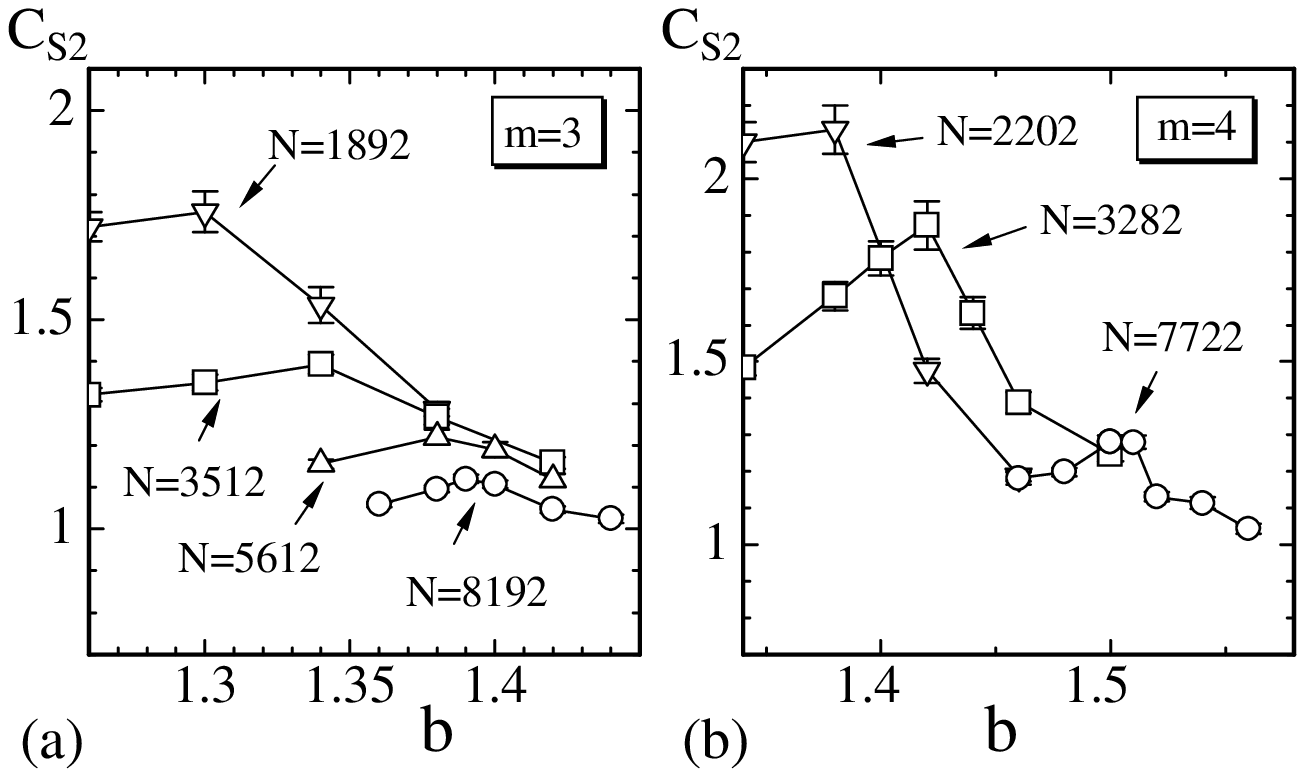}
\caption{The specific heat $C_{S_2}$ vs. $b$ on the surfaces of (a) $m\!=\!3$ and (b) $m\!=\!4$. The error bars are the statistical errors, which were obtained by the binning analysis.} 
\label{fig-11}
\end{figure}
In order to confirm that no surface fluctuation transition occurs, we plot the specific heat defined by
\begin{equation}
C_{S_2} \!=\! {b^2\over N} \langle \; \left( S_2 \!-\! \langle S_2 \rangle\right)^2\rangle
\end{equation}
in Figs.\ref{fig-11}(a) and \ref{fig-11}(b), which respectively correspond to $S_2/N_B$ in  Figs.\ref{fig-10}(a) and \ref{fig-10}(b). We find a peak in each $C_{S_2}$; and this implies that the surface fluctuation becomes large at the peak point. However, the peak values  $C_{S_2}^{\rm max}$ decrease with increasing $N$. This obviously indicates that the surface fluctuation phenomenon is suppressed at large $N$ and, hence, the surface fluctuation is not considered to be a phase transition.  

\section{Summary and conclusion}
A tethered surface model has been investigated on triangulated spherical surfaces with many holes by the canonical Monte Carlo simulations. The ratio of the area of holes to that of the surface including the holes is fixed in the starting configurations, and it is given by $(m-1)/2m$ in the limit of $N\to \infty$, where $m(\!=\!3,4)$ represents the number of partitions of an edge of the icosahedron. As a consequence, the total number of holes remains unchanged when the size $N$ increases under the condition that $m$ is fixed.   

We found a first-order transition of surface collapsing phenomena, which are characterized by a discontinuous change of $X^2$. This transition separates the smooth phase from the collapsed phase. The surfaces are smooth in both phases and, then, the Hausdorff dimension in the smooth state and that in the collapsed state are almost identical, and they are $H\!\simeq\!2$, which is identical to the topological dimension of surfaces. For this reason, it is possible to consider that the transition is observed in biological or artificial membranes, which are not always closed. The model surface can be considered as an open surface, because the size $La$ of holes is comparable (i.e. not negligible compared) to its surface size in the limit of $N\!\to\! \infty$.

Moreover, the collapsing transition is not accompanied by the surface fluctuation transition, which is the one characterized by a discontinuous change of the bending energy $S_2$.  A surface fluctuation phenomenon occurs in the model and, therefore, the specific heat $C_{S_2}$ has a peak at the surface collapsing transition point. However, the peak disappears from $C_{S_2}$ at sufficiently large $N$.

\begin{acknowledgments}
This work is supported in part by a Grant-in-Aid for Scientific Research, No. 18560185.
\end{acknowledgments}



\end{document}